# Experimental evidence of $t_{2g}$ electron-gas Rashba interaction induced by asymmetric orbital hybridization

G. J. Omar[1,†], W. L. Kong[2,†], H. Jani[1], M. S. Li[3], J. Zhou[1], Z. S. Lim[1], S. Prakash[1], S. W. Zeng[1], S. Hooda[4], T. Venkatesan[4], Y. P. Feng[1], S. J. Pennycook[3], L. Shen[2], A. Ariando[1,*]

[1]*Department of Physics, Faculty of Science, National University of Singapore, Singapore 117542, Singapore*

[2]*Department of Mechanical Engineering, National University of Singapore, Singapore 117575, Singapore*

[3]*Department of Materials Science and Engineering, National University of Singapore, Singapore, 117575*

[4]*Department of Electrical and Computer Engineering, National University of Singapore, Singapore 117576, Singapore*

*Corresponding author.
  Email address: ariando@nus.edu.sg
† Authors contributed equally.

**We report the control of Rashba spin-orbit interaction by tuning asymmetric hybridization between Ti-orbitals at the $LaAlO_3/SrTiO_3$ interface. This asymmetric orbital hybridization is modulated by introducing a $LaFeO_3$ layer between $LaAlO_3$ and $SrTiO_3$, which alters the Ti-O lattice polarization and traps interfacial charge carriers, resulting in a large Rashba spin-orbit effect at the interface in the absence of an external bias. This observation is verified through high-resolution electron microscopy, magneto-transport and first-principles calculations. Our results open hitherto unexplored avenues of controlling Rashba interaction to design next-generation spin-orbitronics.**

Spin-orbit coupling (SOC) is an intrinsic property of the material arising due to interaction between its quantum particle's spin $\sigma$ and momentum $k$. This SOC effect has been proposed to be a potential route for the development of energy efficient devices spin transistor[1], spin–orbit qubit, and spin–orbit torque magnetic memory device utilizing spin degree of freedom, popularly known as spin-orbitronic devices[2-4]. Of particular interest is the Rashba SOC[5], which is a relativistic effect associated with inversion symmetry breaking typically found in a low-dimensional systems. Due to this inversion symmetry breaking, an electric field ($E_0$) normal to the interface arises, thereby lifting the spin degeneracy at $k$-points in the Brillouin zone. The Hamiltonian of Rashba system is defined by $H_R = \alpha_R \hat{z} \cdot (\boldsymbol{k} \times \boldsymbol{\sigma})$, where $\alpha_R$ is the Rashba SOC coefficient and $\hat{z}$ is the Rashba unit vector normal to the interface. Non-zero entanglement of $\boldsymbol{k}$ and $\boldsymbol{\sigma}$ alongside broken inversion symmetry produces this effect. It is responsible for the novel emergent phenomena in various condensed matter systems such as graphene, topological insulators, Majorana fermions and cold atoms[6,7].

Complex oxide heterostructures $ABO_3$ based two-dimensional electron system (2DES) are great potential for exploiting the Rashba effect because of their multiple degrees of freedom (charge, spin, orbital, and lattice) are entangled with one another[8,9]. The canonical model of the 2DES at the interface between band insulators $LaAlO_3$ (LAO) and $SrTiO_3$ (STO) have been shown to exhibit a strong Rashba effect with long carrier lifetimes (crucial for low power spintronics[10-13]). Although inversion symmetry is naturally broken at STO interfaces and considerable effort has been devoted in maximizing the Rashba effect[14], exploiting the influence of this effect at zero bias voltage on the electronic states of interest remains elusive.

The build-up symmetry-breaking electric field perpendicular to the interface produces opposite forces on the Ti cations and oxygen anions resulting in lattice polarization. This has been studied by calculating the orbital hybridization of the $t_{2g}$ electrons of Ti atoms in STO at the interface[15]. The Hamiltonian $H$ is defined as $H = H_0 + H_{ASO} + H_z$ where $H_0$ is the intra-orbital hopping (diagonal in the orbital space), $H_{ASO}$ is the on-site atomic SOC, and $H_z$ is the asymmetric inter-orbital hopping[15-18]. These various intra- and inter- orbital perturbed hopping terms between the energy bands influence the energy dispersion and SOC. In particular, the $H_z$ term generates electronic hopping from $d_{xy}$ to $d_{zx}$ along the $y$ direction via $p_x$ and from $d_{xy}$ to $d_{yz}$ along the $x$ direction via $p_y$ in the second-order perturbation producing a Rashba-like SOC effect at the LAO/STO interface. Further discussion is presented in **Supplemental Material SM1**. In the electron momentum $\boldsymbol{k} = (k_x, k_y, 0)$, Pauli matrices $\boldsymbol{\sigma} = (\sigma_x, \sigma_y, \sigma_z)$, and orbital basis $(yz, zx, xy)$, the asymmetric hopping Hamiltonian term $H_z$ takes the form,

$$H_z = \Delta_z \begin{pmatrix} 0 & 0 & ik_x \\ 0 & 0 & ik_y \\ -ik_x & -ik_y & 0 \end{pmatrix} \otimes \sigma^0 \begin{Bmatrix} yz \\ zx \\ xy \end{Bmatrix}, \tag{1}$$

where $\Delta_z = (nt_{pd}^2/\Delta_{pd}) + \gamma_1 t_{pd} E_0/\Delta_{pd}$ is related to the bond angle, n, or ionic polarization/ displacement ($\Delta\delta_{Ti-O}$) along the z direction and induced orbital polarization arising from the additional electric field (with hopping amplitude $E_0\gamma_1$) mediated by the p-d hybridized orbitals with the hopping amplitude $t_{pd}$. Here $\Delta_{pd}$ is the splitting between the O- and Ti-orbitals. This asymmetric hybridization ($\Delta_z$) in the orbital network Ti($3d_{zx}$) – O($2p_x$) – Ti($3d_{xy}$) within the $t_{2g}$ manifold directly leads to Rashba splitting. This $\Delta_z$ is sensitive to local lattice polarization at the LAO/STO interface. The $\Delta_z$ is a layer-dependent parameter, having a maximum value at the interface and decreasing rapidly at the deeper layers resulting into a layer dependent ionic displacement. Larger ionic displacement induces more built-in electric field and thus results in enhanced Rashba SOC. Notably, the Rashba parameter is directly proportional to the asymmetric hopping term $\Delta_z$, which is directly related to the induced orbital polarization and the ionic displacement.

In this work, we provide first experimental evidence for enhancement of Rashba SOC (at zero bias) at LAO/STO interfaces through asymmetric orbital hybridization. To examine the pronounced contribution of the antisymmetric hopping, LAO and STO interface is modulated with a LaFeO$_3$ (LFO) buffer layer with thicknesses $d$ = 0 to 6-unit cells (uc). We demonstrate enhancement of the Rashba SOC by introducing x uc of LFO layer. Samples were fabricated using pulsed laser deposition (PLD) system. A real time reflection high-energy electron diffraction (RHEED) method was utilized to monitor the layer-by-layer growth mode and to control the layer thickness of the thin films. (**Supplemental Material SM6**). The atomic structure of the interfaces was characterized using a cross-section high-angle annular dark-field (HAADF) mode of the scanning transmission electron microscopy (STEM). STEM images confirm coherent and epitaxial growth with atomically sharp interfaces (**Supplemental Material SM5**). As discussed previously, the ionic displacement ($\Delta\delta_{Ti-O}$) is associated with an internal electric field $E_0$ and is directly responsible for the Rashba interaction parameters. We, therefore, calculated the increase in the distortion of the TiO$_2$ plane in LAO/STO interfaces with and without LFO layer using First-principles calculations with Vienna ab-initio simulations package

(VASP). We further validated the SOC enhancement through magnetoconductance transport and its weak antilocalization fitting.

A LAO/LFO/STO interface showing a Ti-O-Ti ionic displacement ($\Delta\delta_{Ti-O}$) along the [001] axis for the topmost TiO$_2$ plane is shown in **Fig. 1(a)**. The ionic displacements derived from the first-principles calculations ($\Delta\delta_{Ti-O}$) of the TiO$_2$ layer at different depths from the interface for $d = 0$ and 4 uc are given in **Fig. 1(b)**. The lattice distortion is calculated from the displacement of the ions along the [001] axis relative to the center of the Ti- and O-sites. We can see that for the heterostructure with a LFO layer, the first STO layer near the interface has a pronounced ionic displacement ($\Delta\delta_{Ti-O}$) compared to that without the LFO layer. We verify this lattice displacement experimentally using annular bright-field (ABF) mode in the STEM for the $d = 4$ uc sample (**Fig. 1(c)**). Lattice displacement value (shift of O-anions and Ti-cations from their ideal lattice positions in the antiparallel direction) averaged over each layer with the error bars is plotted in **Fig. 1(e)**. We observed a significantly larger displacement of Ti-cations in the topmost STO layer as compared to the previous reports of LAO/STO heterostructure without the LFO layer[19]. This supports our hypothesis of increased effective electric field at the interface with the incorporation of the modulation layer (LFO), thereby resulting in the pronounced enhancement of the Rashba SOC. Our results can be explained by comparing the asymmetric hybridization ($\Delta_z$) and ionic displacement ($\Delta\delta_{Ti-O}$) in the orbital network Ti($3d_{zx}$) – O($2p_x$) – Ti($3d_{xy}$) of the interfacial TiO$_2$ layer with and without LFO layer as shown schematically in **Fig. 1(a)**.

Further, we show the sheet resistance ($R_{xx}$) as a function of temperature between 3-300 K for LAO/LFO/STO heterostructures in **Fig. 2(a)**. Increasing $d$ leads to a suppression of the charge transfer and the 2DES carrier density, resulting in enhancement of the longitudinal resistivity. Eventually, a metal-to-insulator transition is observed at $d = 6$ uc (yellow shaded area in **Fig. 2(a)**), as we have reported earlier[20]. As the LFO layer is also polar, the relatively smaller internal electric field of the LFO per unit cell, however, would require ~50 uc of LFO to have enough electric field to induce a charge transfer[11,21]. Hence, all our LFO/STO interfaces are indeed always insulating[20]. **Fig. 2(b)** shows the sheet carrier densities ($n_S$) of our LAO/LFO/STO heterostructures, which are typically lower than the carrier density ($1.68 \pm 0.18 \times 10^{13}$ cm$^{-2}$) of the 2DES at the LAO/STO interface[22]. The low carrier density indicates that only one of the Ti $3d$ orbitals is being occupied, that is the predominantly lowest Ti $d_{xy}$ band. The linear Hall resistance up to 9 T and its multicarrier band model fitting further supports predominant single-band occupancy. The electron diffusion constant, calculated as $D = \frac{1}{2}v_F^2 \tau_e$ for this singly occupied $d_{xy}$ orbital, where Fermi velocity $v_F = \hbar k_F/m^*$, Fermi momentum $k_F = \sqrt{2\pi n_S}$, and $m^*$ is the effective mass, decreases with increasing $d$ (**Fig. 2(c)**).

Experimentally, Rashba SOC was typically calculated from low-temperature magnetotransport using a quantum correction associated with the weak antilocalization (WAL)[23-25] and tuning of the Rashba SOC has also been demonstrated using electrostatic fields[26-31]. We, therefore, measured magnetoresistance (MR) on our 2DES samples using a dc four-probe technique in a Hall bar configuration (**Supplemental Material SM7**). The observed MR is typically positive and increases quadratically in high fields. As the temperature is decreased below 6 K, a cusp feature starts to emerge around zero fields. Moreover, the MR shows no dependency on the applied current (**Supplemental Material SM7**). The positive MR with a cusp has been attributed to WAL and the presence of a strong SOC, similar to that reported in thin metallic films and semiconductor heterostructures[23,32-36]. A similar enhancement of the cusp feature has been observed in the LAO/STO heterostructure[26,28] under the application of large electric fields of up to 100 V. Even though LFO reduces the carrier density and shifts the Fermi level below Lifshitz transition, we still achieved SOC enhancement. Conventionally, in the electric-field induced SOC, the enhancement is always accompanied by the increase in the carrier density (Fermi level shift across energy bands, Lifshitz transition)[22]. On the contrary, our SOC enhancement is accompanied by a reduction in the carrier density in the absence of an external electric field.

To further understand the spin relaxation mechanism we determine the first-order quantum correction to the magnetoconductance, $\Delta\sigma_S/\sigma_0 \equiv (\sigma_S(H)/\sigma_0 - 1)$, by fitting it to the Maekawa-Fukuyama (MF) model[37]. The fitting model details are described in Supplemental Material SM3. Our thickness-dependent magnetoconductance data fits well with this theory, considering two independent parameters inelastic ($\tau_i$) and spin-orbit relaxation scattering time ($\tau_{SO}$). This results in non-trivial evolution of effective inelastic ($H_i$) and spin-orbit field ($H_{SO}$), and elastic ($\tau_e$), inelastic ($\tau_i$), and spin-orbit relaxation scattering time ($\tau_{SO}$). In particular, the effective spin-orbit field ($H_{SO}$) (seen by the electrons moving relativistically under the influence of the $E_0$) increases by an order of magnitude with $d$ while $H_i$ remains nearly constant as shown in **Fig. 3(a)**. Further, while $\tau_i$ is almost constant at 2 ps for all samples, the $\tau_{SO}$ decreases by one order of magnitude from 0.3 ps for the as-grown LAO-STO to 0.03 ps for the samples with a buffer layer. These results are consistent with the expectation from the MF model in the strong SOC regime where $\tau_{SO} < \tau_i$. We also find that the $\tau_{SO}$ is inversely proportional to the $\tau_e$ over the range of the $d$ values in our experiment. This is a clear sign of the dominating DP spin relaxation mechanism[38,39] (via Rashba SOC) wherein the electron spins precess around the $H_{SO}$ field with a corresponding Larmor frequency, $\Omega(\boldsymbol{k})$ (**Supplemental Material SM7**). A similar spin relaxation mechanism was also reported in the gate tunable LAO/STO system[26]. We further determine the Rashba SOC coefficient $\alpha_R$ from the relation $\alpha_R = \hbar^2/(2m^{*2}\sqrt{\tau_{SO}D})$, where $m^* = 3m_e$ is electron's effective mass[40]. Here, the Rashba effect lifts the double degeneracy of the $3d$ orbitals and results in a spin splitting ($\Delta_R = 2\alpha_R k_F$) of a few meV. **Fig. 3(b)** demonstrates the enhancement of the SOC and the spin splitting due to the introduction of the carrier modulation layer. It is worth noting here that although the modulation layer and the electrostatic gating both can result in a SOC enhancement, the dependence of the $\alpha_R$ on the $n_S$ is the opposite in the two cases. In the electrostatic gating, a positive back-gate voltage leads to an increase in the $n_S$ as expected and thus results in the enhancement of the $\alpha_R$. Whereas in our system, increasing the LFO thickness leads to a decrease in $n_S$, yet $\alpha_R$ increase significantly. Considering lower $n_S$ (predominantly $d_{xy}$ orbitals occupied), the results are following the expectations of the $t_{2g}$ Rashba theory as discussed previously. To compare the strength of the Rashba SOC of the heterostructures with and without the LFO layers, the Rashba coefficients at the $\Gamma$ point of the lowest $d_{xy}$ band were calculated (**Fig. 3(c)**). Notably, the spin momentum locking is direct evidence of the presence of Rashba SOC in the STO band structure. The band structures of the LAO/LFO(4 uc)/STO and the LAO/STO including SOC are calculated along the $\Gamma - M$ direction in the $k$ space (**Supplemental Material SM4**). The interfacial symmetry breaking lifts the degeneracy of the Ti $t_{2g}$ bands, resulting in the splitting between the $d_{xy}$ and the $d_{zx}, d_{yz}$ bands. The conduction band minimum at the Γ point is mainly formed by the $d_{xy}$ orbitals of the interface Ti-sites. For the standard Rashba SOC, the spin splitting Δ is linearly dependent on $k$, where the Rashba coefficient $α_R$ can be defined as $Δ/2|k|$[41]. The Rashba coefficients of the heterostructures with and without LFO layers are 4.3 meV Å and 0.6 meV Å, respectively (**Fig. 3(d)**). This result provides strong theoretical support to the experimentally observed enhancement of the Rashba coefficient due to the incorporation of the LFO layer.

To further understand the origin of the SOC tuning, we have performed EELS in a cross-sectional STEM. The EELS data of LAO/LFO($d$ uc)/STO are simultaneous acquired with ADF images where $d$ = 0, 2, and 4 uc. Layer resolved Fe-$L$ and Ti-$L$ EELS at different FeO$_2$ and TiO$_2$ planes in LFO and STO layers are analysed and formation of $Fe^{2+}$ and $Ti^{3+}$ concentration and distribution for the $d$ = 0, 2, and 4 uc samples are presented. A summary of the atomic layer resolved charge distribution for the $d$ = 0, 2 and 4 uc samples is shown in **Fig. 4(a)-(c)**, where the reduced Ti and Fe fractions are defined as $Δ_{Ti}$ = $Ti^{3+}$/($Ti^{3+}$+$Ti^{4+}$) and $Δ_{Fe}$ = $Fe^{2+}$/($Fe^{2+}$+$Fe^{3+}$), respectively. Clearly, the electron trapping in the LFO results in less free carriers in the STO layer compared to the control LAO/STO interface. From these results, we extract the integrated distribution ($A_d$), width ($w_d$) and peak ($Δ_{Ti}^1$) of the $Ti^{3+}$ in the STO layers as a function of $d$ (**Fig. 4(d)-(f)**, respectively). Both $Δ_{Ti}^1$ and $A_d$ decrease with $d$, suggesting that the $Ti^{3+}$ decreases with the increase of the LFO thickness. This result agrees with our Hall data (**Fig. 3(d)**), which show a suppression of total carrier density ($n_S$). The $w_d$, which is related to the spatial width of the electronic distribution, also decreases with $d$, suggesting that the spatial confinement of 2DES in STO increases in the presence of the LFO layers. These results provide further insight into the role of LFO layers in tuning the inversion asymmetry and enhancing the effective electric field experienced by the 2DES[42-44].

In conclusion, our work highlights the fundamental role of asymmetric orbital hybridization mediated via lattice polarization in achieving the observed enhancement of Rashba SOC. First-principles calculations and ABF imaging have provided clear evidence of increased lattice displacement in the $TiO_2$ planes by incorporation of LFO layer. This strategy for tuning and controlling Rashba SOC through lattice polarization is particularly promising as it can be integrated directly in functional devices[45] for efficient spin-to-charge conversion[46-49]. Moreover, it can lead to the discovery of various exotic properties, such as spiral magnetism[50], topological superconductivity[51], and intrinsic spin Hall effect[52].

**Acknowledgments**

This research is supported by the Agency for Science, Technology and Research (A*STAR) under its Advanced Manufacturing and Engineering (AME) Individual Research Grant (IRG) (A2083c0054). The authors also acknowledge the Singapore National Research Foundation (NRF) under the Competitive Research Programs (CRP Award No. NRF-CRP15-2015-01) for a partial support and the Centre for Advanced 2D Materials at National University of Singapore for providing computing resources.

**Competing interests**

The authors declare no competing interests.

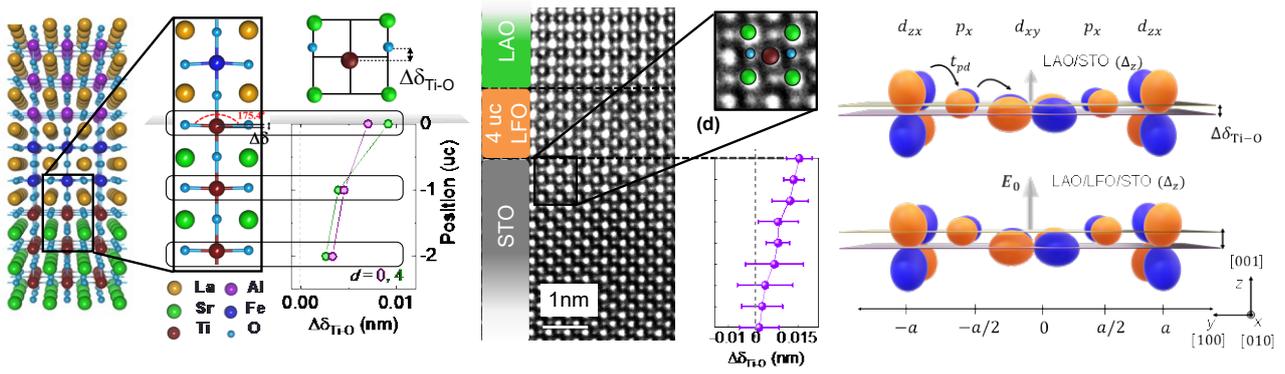

**Fig. 1** (a) Schematic figure of LAO/LFO/STO in the (001) crystallographic direction. Zoomed-in ionic displacement patterns ($\Delta\delta$) in the unit-cell structures calculated from first-principles calculations. (b) Ti-O-Ti ionic displacement ($\Delta\delta_{Ti-O}$) along the [001] axis for with ($d = 4$) and without ($d = 0$) LFO layer. (c) ABF-STEM results for LAO/LFO(4uc)/STO heterostructure. (d) $\Delta\delta_{Ti-O}$ is calculated from the displacement (with error bar) along the [001] axis between the center position of Ti-site cation and the O-site anions in ABF-STEM. (e) Orbital bonding network between Ti $d_{zx}$ and $d_{xy}$ asymmetric orbitals on neighbouring metal atoms through $p_x$ orbitals along the $y$ axis with build-in electric field. Comparison of displacement of the Ti cation ($d_{xy}, d_{zx}$ orbital) (top plane) and oxygen ($p_x$) (bottom plane) sublattices in large built-in electric field for with and with LFO layer. The schematic positive and negative lobes of the orbital functions are represented in orange and blue, respectively.

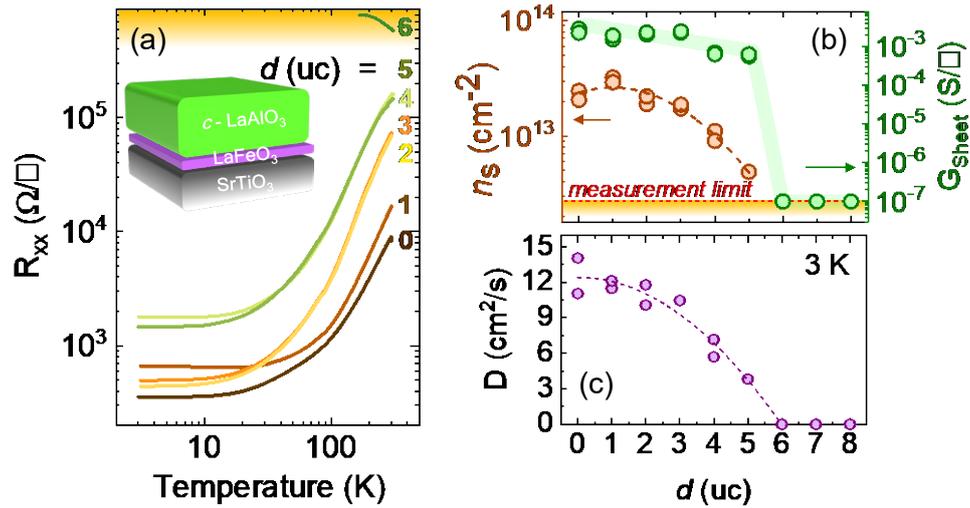

**Fig. 2** (a) The temperature dependence of the transverse sheet resistance ($R_{xx}$) for various $d$ in 2DES heterostructure. With the characteristic length above $d$ = 6 uc of LFO in heterostructure becomes insulating (yellow shaded area). (b), Left axis, orange circles: LFO-thickness-dependent carrier density ($n_S$) and right axis, green circles: sheet conductance ($G_{Sheet}$) at 3 K. (c), Thickness dependent modulation of the diffusion coefficient $D$ at 3 K. The dotted lines are guides to the eye.

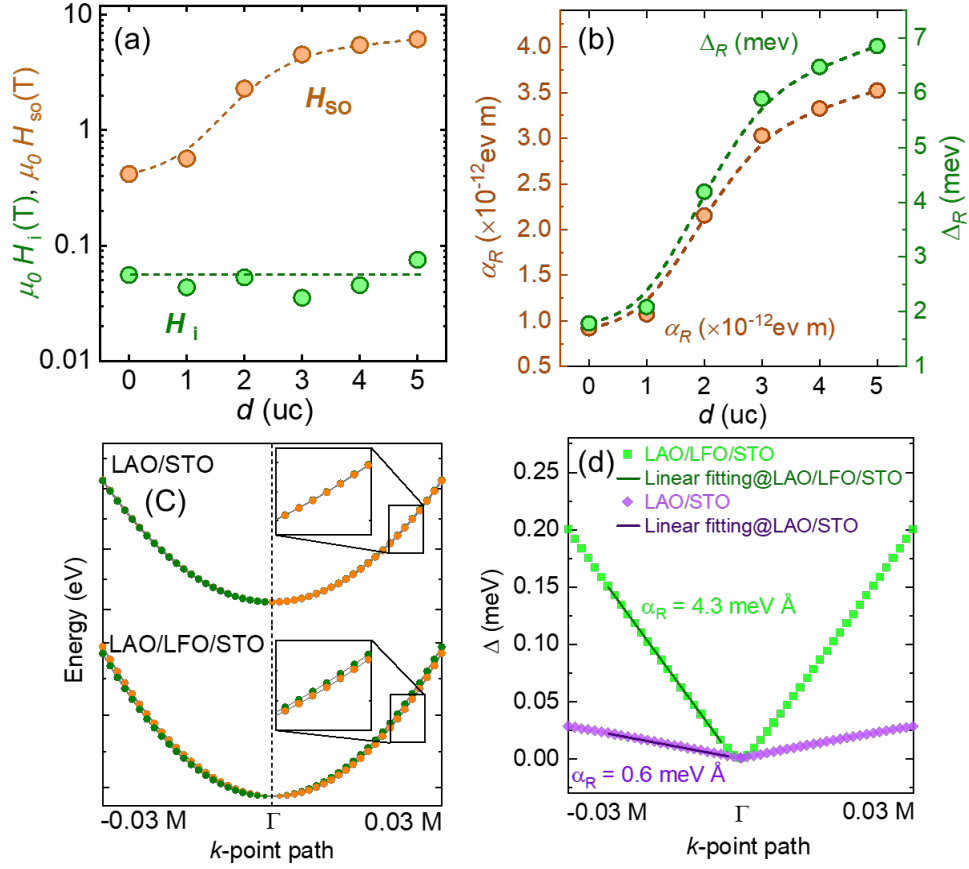

**Fig. 3** (a) LFO thickness ($d$) dependence of the fitting parameters $H_i$ (green circles) and $H_{SO}$ (orange circles). (b) Left axis, orange circles: Rashba SOC Coefficient, $\alpha_R$ and right axis, green circles: Rashba spin splitting, $\Delta_R$ as a function of $d$. The dotted lines are guides to the eye for all plots. (c) The closer zoomed-in DFT bands around the Γ point (zoomed-in Rashba spin splitting at a smaller range of $k$-point path, inset figures). The orange and dark green dots denote spin components with opposite directions (oriented along the $y$ axis) lying perpendicular to the $k$-vector (along the $x$ axis). The size of each dot denotes the magnitude of the corresponding spin component. (d) Rashba spin splitting for LAO/STO and LAO/LFO/STO shows the linear momentum dependence.

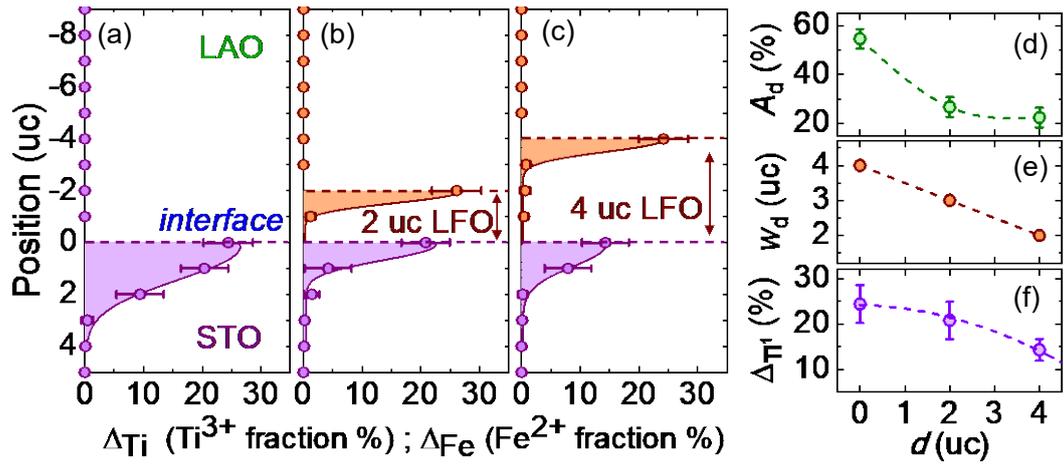

**Fig. 4** (a)-(c), A summary of the atomic layer resolved charge distributions (with error bar) in % defined as the $Ti^{3+}$ fraction $\Delta_{Ti} = Ti^{3+}/(Ti^{3+}+Ti^{4+})$ and $Fe^{2+}$ fraction $\Delta_{Fe} = Fe^{2+}/(Fe^{2+}+Fe^{3+})$ for $d$ = (a) 0, (b) 2, and (c) 4 uc. The lines are a guide to the eye. (d)-(f), The integrated distribution ($A_d$), spatial width of the electronic distribution ($w_d$) and peak ($\Delta_{Ti}^1$) of the $Ti^{3+}$ species in the first interfacial layer of STO, as a function of $d$ = 0, 2, and 4 uc. The dotted lines are a guide to the eye.

# Supplemental Material

# Experimental evidence of $t_{2g}$ electron-gas Rashba interaction induced by asymmetric orbital hybridization

G. J. Omar[1,†], W. L. Kong[1,†], H. Jani[1], M. S. Li[2], J. Zhou[1], Z. S. Lim[1], S. Prakash[1], S. W. Zeng[1], S. Hooda[3], T. Venkatesan[3], Y. P. Feng[1], S. J. Pennycook[2], L. Shen[4], A. Ariando[1,*]

[1]*Department of Physics, Faculty of Science, National University of Singapore, Singapore 117542, Singapore*

[2]*Department of Materials Science and Engineering, National University of Singapore, Singapore 117575, Singapore*

[3]*Department of Electrical and Computer Engineering, National University of Singapore, Singapore 117576, Singapore*

[4]*Department of Mechanical Engineering, National University of Singapore, Singapore, 117575*

**CONTENTS**

## SM1. Theory of $t_{2g}$ electron-gas Rashba interactions.

Two-dimensional electron gas where the confined electrons of the conduction band are in Ti-$3d$ orbitals, splits into $t_{2g}(3d_{xy}, 3d_{xz,yz})$ and $e_g(3d_{x^2-y^2}, 3d_{z^2})$ orbitals due to crystal field. The interfacial symmetry breaking lifts the degeneracy of the Ti $t_{2g}$ bands, resulting in the splitting between the $d_{xy}$ and the $d_{zx}, d_{yz}$ bands. The Hamiltonian for these $t_{2g}$ bands of Ti cations at the top layer of STO is defined as $H = H_0 + H_{ASO} + H_z$ where $H_0 = H_{atomic} + \Delta U$, $H_{atomic}$ defined as atomic orbital Hamiltonian and $\Delta U$ is the terms related to the hopping terms between nearest-neighbor orbital. $H_{ASO} + H_z$ are the perturbed hopping Hamiltonian terms. Here, $H_{ASO}$ is on-site atomic SOC and $H_z$ is antisymmetric inter-orbital hopping. In the simplistic electronic structure, $t_{2g}$ $xy$ orbitals on one B site can hop only along the $y$ or $x$ direction through an intermediate $p_x$ or $p_y$ orbital to an $xy$ orbital on the B site of a neighbouring cubic cell [1].

$$H_0 = \begin{pmatrix} \frac{k_x^2}{2m_h} + \frac{k_y^2}{2m_l} & 0 & 0 \\ 0 & \frac{k_x^2}{2m_h} + \frac{k_y^2}{2m_l} & 0 \\ 0 & 0 & \frac{k_x^2}{2m_h} + \frac{k_y^2}{2m_l} - \Delta_E \end{pmatrix} \otimes \sigma^0 \begin{Bmatrix} yz \\ zx \\ xy \end{Bmatrix}, \quad (S1)$$

where, $\Delta_E$ is the energy difference between the $d_{xy}$ band and the $d_{xz}, d_{yz}$ orbitals due to the transverse confinement along the $z$ direction. $H_{ASO}$ is atomic spin-orbit coupling projected to $t_{2g}$ orbital bands,

$$H_{ASO} = \Delta_{ASO} \begin{pmatrix} 0 & i\sigma_z & -i\sigma_y \\ -i\sigma_z & 0 & i\sigma_x \\ i\sigma_y & -i\sigma_x & 0 \end{pmatrix} \begin{Bmatrix} yz \\ zx \\ xy \end{Bmatrix}, \quad (S2a)$$

$$= \begin{pmatrix} 0 & 0 & i & 0 & 0 & -1 \\ 0 & 0 & 0 & -i & 1 & 0 \\ -i & 0 & 0 & 0 & 0 & i \\ 0 & i & 0 & 0 & i & 0 \\ 0 & 1 & 0 & -i & 0 & 0 \\ -1 & 0 & -i & 0 & 0 & 0 \end{pmatrix} \begin{Bmatrix} yz, \uparrow \\ yz, \downarrow \\ zx, \uparrow \\ zx, \downarrow \\ xy, \uparrow \\ xy, \downarrow \end{Bmatrix}, \quad (S2b)$$

where, $\Delta_{ASO}$ is atomic spin-orbit mixing term. $H_z$ is antisymmetric interorbital nearest-neighbour hopping, a layer dependent term, induced by polar lattice displacement due to the electric field ($E_0$) from broken inversion symmetry.

$$H_z = \Delta_z \begin{pmatrix} 0 & 0 & ik_x \\ 0 & 0 & ik_y \\ -ik_x & -ik_y & 0 \end{pmatrix} \otimes \sigma^0 \begin{Bmatrix} yz \\ zx \\ xy \end{Bmatrix}, \quad (S3)$$

where, $\Delta_z$ generates hopping terms from $d_{xy}$ to $d_{xz}$ only in the $y$ direction through $p_x$ and from $d_{xy}$ to $d_{yz}$ only in the $x$ direction through $p_y$. The perturbation arises from additional potential $-eE_0z$, which induce hopping amplitude as $E_0\gamma_1$, where $\gamma_1 = \langle d_{zx}, \vec{R} = 0| -ez | p_x, \vec{R} = \frac{a}{2\hat{y}} \rangle$

In the second-order perturbation, Ti($3d_{xy}$) – O($2p_x$) – Ti($3d_{xy}$) hopping described by the effective transfer integral $t_{1,2}$ as below,

$$t_{1,2} = \frac{\langle d_{xy}, \vec{R} = 0|H_0|p_x, \vec{R} = \frac{a}{2}\hat{y}\rangle \langle p_x, \vec{R} = \frac{a}{2}\hat{y}|H_0|d_{xy}, \vec{R} = a\hat{y}\rangle}{\Delta_{pd}} \quad (S4)$$

$$t_{1,2} = \frac{t_{pd}^2}{\Delta_{pd}} \quad (S5)$$

where, $t_{pd}$ is the orbital hopping amplitude of the $p-d$ hybridization and $\Delta_{pd}$ is the spitting between the oxygen $p$ and Ti $t_{2g}$ $3d_{xy}$ orbital. In the second order perturbation, Ti($3d_{zx}$) – O($2p_x$) – Ti($3d_{xy}$) hopping is defined as (Figure S1),

$$\Delta_z = \frac{\left\langle d_{zx}, \vec{R}=0 \right| -eE_0 z \left| p_x, \vec{R}=\frac{a}{2}\hat{y} \right\rangle \left\langle p_x, \vec{R}=\frac{a}{2}\hat{y} \right| H_0 \left| d_{xy}, \vec{R}=a\hat{y} \right\rangle}{\Delta_{pd}} \tag{S6}$$

$$\Delta_z = \frac{E_0 \gamma_1 t_{pd}}{\Delta_{pd}} \tag{S7}$$

Finally, the expectation value of second-order perturbed Hamiltonian,

$\langle k, d_{xy}\sigma' | H^{(2)} | k, d_{xy}, \sigma \rangle$

$$= \sum_{k',\sigma''} \frac{\langle k, d_{xy}\sigma' | H_z | k', d_{xz}, \sigma'' \rangle \langle k', d_{xz}\sigma'' | H_{ASO} | k, d_{xy}, \sigma \rangle}{E_{d_{xz}}(k') - E_{d_{xy}}(k)}$$

$$+ \frac{\langle k, d_{xy}\sigma' | H_z | k', d_{yz}, \sigma'' \rangle \langle k', d_{yz}\sigma'' | H_{ASO} | k, d_{xy}, \sigma \rangle}{E_{d_{yz}}(k') - E_{d_{xy}}(k)}$$

$$= \epsilon_{xy}(\vec{k}) - \frac{\Delta_z \Delta_{ASO}}{\Delta_E}[\langle \sigma_y \rangle k_x a - \langle \sigma_x \rangle k_y a] \tag{S8}$$

$$= \epsilon_{xy}(\vec{k}) - \alpha_R (\vec{k} \times \vec{\sigma}) \cdot \hat{z}$$

where, $\alpha_R \sim \frac{\Delta_z \Delta_{ASO}}{\Delta_E} a$, defined as conventional Rashba spin-orbit coupling, valid for small $k$ and for the $xy$ band. Notably, the Rashba parameter is directly proportional to the asymmetric hopping term $\Delta_z$, which is directly related to the induced orbital polarization and the atomic displacement.

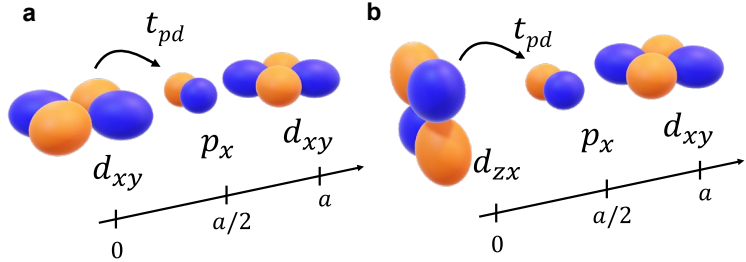

**Figure S1.** Schematic figure of the Ti metal and oxygen orbitals at the STO interface resulting in **a**, Ti($3d_{xy}$) – O($2p_x$) – Ti($3d_{xy}$) and **b**, Ti($3d_{zx}$) – O($2p_x$) – Ti($3d_{xy}$) hopping with the hopping amplitude $t_{pd}$.

## SM2. Sample preparation.

500 $\mu$m-thick SrTiO$_3$ (STO) (001) substrates were treated with HF solution and annealed at temperature $T = 950$ °C to obtain atomic steps and TiO$_2$-terminated surfaces. The STO substrates were patterned in a Hall bar and lateral gate electrode geometry by using conventional photolithography and amorphous highly insulating AlN films were deposited as a hard mask wall as shown in the inset of Figure S10. Pulsed laser deposition (PLD) method was used for sample preparation and the layer-by-layer growth was monitored by *in-situ* reflection high energy electron diffraction (RHEED). The LaFeO$_3$ (LFO) layer thickness ($d$ uc) was changed from 0 to 8 uc, while the LaAlO$_3$ (LAO) layer thickness was kept fixed at 8 uc. A nanosecond KrF 248nm laser was used with fluence of 1.5 Jcm$^{-2}$ and a repetition rate of 2 Hz to grow all samples. We used the growth parameters, with temperature $T = 750$ °C, oxygen pressure $P_{O2}$ = 10 mTorr to grow LFO layers and $T = 740$ °C, $P_{O2}$ = 0.5 mTorr to grow LAO layer on top.

## SM3. The fit of the magnetoconductance curves using the theoretical models.

The Iordanskii, Lyanda-Geller, and Pikus (ILP) [2] is the most developed and widely used theoretical model reported in the literature for the comprehensive description of WL/WAL phenomena. This theory considers both the $k$-linear and $k$-cubic components of Rashba interactions, where $k$ is the wave vector of the electronic carriers. Here, the quantum correction of the magnetoconductance due to the SOC is given as,

$$\frac{\Delta\sigma_S(H)}{\sigma_0} = -\frac{1}{a_0} - \frac{2a_0 + 1 + H_{SO}/H}{a_1(a_0 + H_{SO}/H) - 2H'_{SO}/H}$$
$$+ \sum_{n=1}^{\infty}\left[\frac{3}{n} - \frac{3a_n^2 + 2a_n H_{SO}/H - 1 - 2(n+1)H'_{SO}/H}{(a_n + H_{SO}/H)a_{n-1}a_{n+1} - 2H'_{SO}/H[(2n+1)a_n - 1]}\right]$$
$$- 2\ln\frac{H_{tr}}{H} - \psi\left(\frac{1}{2} + \frac{H_i}{H}\right) - 3c \tag{S9}$$

where $a_n = n + \frac{1}{2} + \frac{H_i}{H} + \frac{H_{SO}}{H}$, $H_i = \frac{\hbar}{4eD\tau_i}$, $H_{tr} = \frac{\hbar}{4eD\tau_1}$, $H_{SO} = \frac{\hbar}{4eD}(2\Omega_1^2\tau_1 + 2\Omega_3^2\tau_3)$, $H'_{SO} = \frac{\hbar}{4eD}2\Omega_1^2\tau_1$, $\frac{1}{\tau_m} = \int(1 - \cos m\theta)W(\theta)d\theta$, $\psi(1 + z) = -c + \sum_{n=1}^{\infty}\frac{z}{n(n+z)}$, $\psi$ is the digamma function, $c$ is the Euler constant, $H_{tr}$ is the elastic scattering field of electrons. $\tau_1$ is the elastic scattering time which can be calculated from the carrier concentration $n_S$ and mobility $\mu_S$. $m = 1$ or $3$ and $W(\theta)$ the probability of scattering by a scattering angle $\theta$. $\sigma_0 = e^2/\pi\hbar$ is a universal value of quantum conductance.

As we mentioned, Equation S9 includes both the $k$-linear $\vec{\Omega}_1$ and $k$-cubic $\vec{\Omega}_3$ spin-orbit terms, where vector $\vec{\Omega}$ is the precession of spins and its direction defines the axis of the precession. It is also related to the spin-splitting parameter $\Delta_R$ through the relation, $\Delta_R = \hbar|\vec{\Omega}|$. The ILP theory includes both field-dependent and -independent terms, where independent- or zero-field contribution also needs to be subtracted according to the experiments. Moreover, omitting the $H'_{SO}$ contribution from the Equation S9 can easily lead to the well-known Hikami-Larkin-Nagaoka (HLN) theory [3], where only the $k$-cubic spin-orbit is present. Later, Maekawa-Fukuyama (MF) model [4] was developed with the inclusion of the Zeeman effect. This model is based on the contribution of the D'yakonov-Perel spin precession mechanism [5], and therefore, it is used to describe the Rashba Coefficients at our oxide interfaces. Generally, the Zeeman effect is suppressed due to the strong SOC with the perpendicular magnetic field.

$$\frac{\Delta\sigma_S(H)}{\sigma_0} = \psi\left(\frac{H}{H_i + H_{so}}\right) + \frac{1}{2\sqrt{1-\gamma^2}}\psi\left(\frac{H}{H_i + H_{so}(1+\sqrt{1-\gamma 2})}\right)$$
$$- \frac{1}{2\sqrt{1-\gamma^2}}\psi\left(\frac{H}{H_i + H_{so}(1-\sqrt{1-\gamma 2})}\right) \quad (S10)$$

Additionally, the contribution from the classical orbital term is added to the Equation S10,

$$\Delta\sigma_{orb} = A_k \frac{H_0}{\sigma_0}\left(\frac{H^2}{1+cH^2}\right) \quad (S11)$$

which becomes relevant at high magnetic fields. In order to estimate the classical orbital term, we first used Equation S11 to fit the high field region (between 4 and 9 T) of our data and obtained an estimation of the parameter $A_k$. Then, we performed a complete fit of the magnetoconductance curve, using the MF formula in Equation S10, with the addition of the S11 term. This process yielded good fits and parameters with high accuracy, for all values of $d$.

## SM4. First-principles calculations of the heterostructures.

First-principles calculations were performed using density-functional theory (DFT) based Vienna *ab initio* simulation package (VASP). Perdew-Burke-Ernzerhof (PBE) approximation was used in the calculations to describe the electron exchange-correlations [6-8]. Interactions between electrons and ions were described using the projector augmented wave (PAW) potentials [9]. PBE+ Hubbard $U$ of 2.5 eV and 8 eV were applied to Fe $d$ orbitals and the unoccupied La $f$ orbitals, respectively, to include the on-site Coulomb interactions [10]. The electronic wave functions were expanded using the plane-wave basis set with a cut-off energy of 500 eV. A $\sqrt{2} \times \sqrt{2} \times 1$ supercell of the vacuum/LAO/LFO/STO slab model was used to simulate the G-type antiferromagnetism of LFO layers in the LAO/LFO/STO heterostructure. For comparison, $\sqrt{2} \times \sqrt{2} \times 1$ supercell of the vacuum/LAO/STO slab model was used to simulate the LAO/STO heterostructure. The thicknesses of LAO, LFO, and STO were set to 6, 4, and 4 uc, respectively, and the vacuum layers were set to about 20 Å to minimize electronic interactions between periodic images. Γ-centered $k$-point grids for sampling the first Brillouin zone were set to $8 \times 8 \times 1$ for the $\sqrt{2} \times \sqrt{2} \times 1$ supercells of both vacuum/LAO/STO and vacuum/LAO/LFO/STO slabs. All atoms except those of the bottom STO layer were relaxed until the force on each atom is less than 0.02 eV/Å.

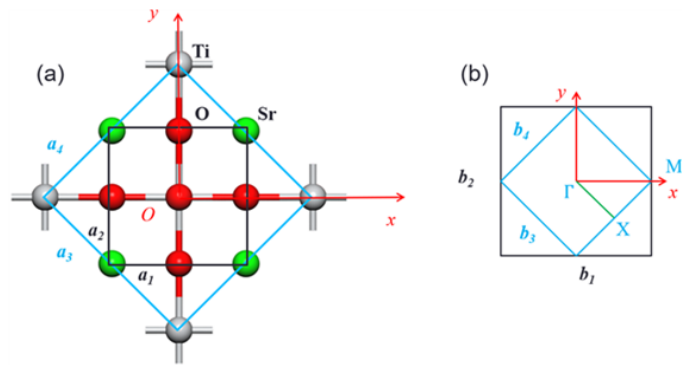

**Figure S2.** (a) Schematic top views of the unit cell (defined by solid black lines) and $\sqrt{2} \times \sqrt{2} \times 1$ supercell (defined by solid blue lines) of the LAO/LFO/STO (LAO/STO) heterostructure, where $a_1$ and $a_2$ denote in-plane lattice parameters of the unit cell while $a_3$ and $a_4$ denote in-plane lattice parameters of the supercell. (b) Schematic top views of first Brillouin zones of the unit cell (defined by solid black lines) and $\sqrt{2} \times \sqrt{2} \times 1$ supercell (defined by solid blue lines) of the LAO/LFO/STO (LAO/STO) heterostructure, where $b_1$, $b_2$, $b_3$, and $b_4$ denote reciprocal lattice parameters corresponding to $a_1$, $a_2$, $a_3$, and $a_4$ in (a), respectively. The $x$-direction of the supercell in (a) is along the Ti-O bond while it is along $\Gamma - M$ of the reciprocal lattice of the supercell in (b).

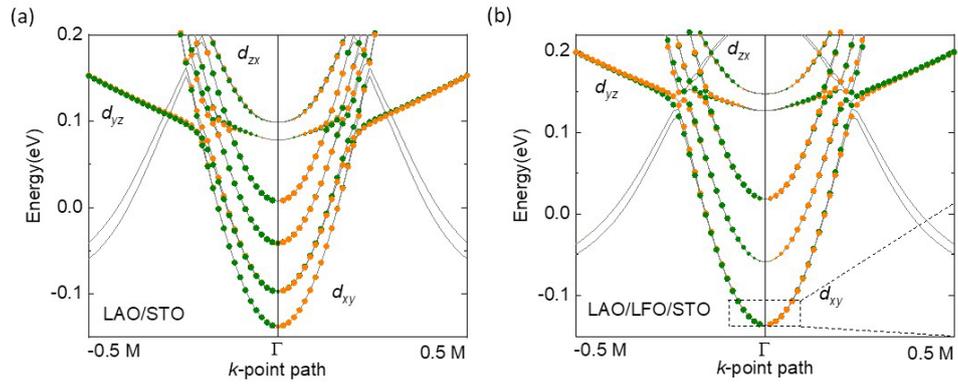

**Figure S3.** Band structure of $t_{2g}$ $d$-orbitals in the STO layer of the LAO/STO and LAO/LFO/STO interface. The orange and dark green dots denote spin components with opposite directions (oriented along the $y$ axis) lying perpendicular to the $k$-vector (in the $\Gamma - M$) (along the $x$ axis). The size of each dot denotes the magnitude of the corresponding spin component. Oxygen $p$ bands are in grey colour.

## SM5. STEM Imaging and EELS analysis

TEM samples were prepared by a FIB (focused ion beam) system (FEI Versa 3D) with 30 kV Ga ions, followed by a low-voltage (i.e., 2 kV) cleaning step. STEM imaging was performed by a JEM-ARM200F (JEOL) microscope (operated at 200 kV), equipped with an ASCOR aberration corrector, a cold-field emission gun and a Gatan Quantum ER spectrometer. The EELS results were acquired using a collection angle of 100 mrad with the energy dispersion of 0.25 and 0.1 eV per channel for the elemental mapping and energy loss near edge structure (ELNES), respectively. The EELS maps were subject to noise reduction by a principal component analysis (PCA) filter.

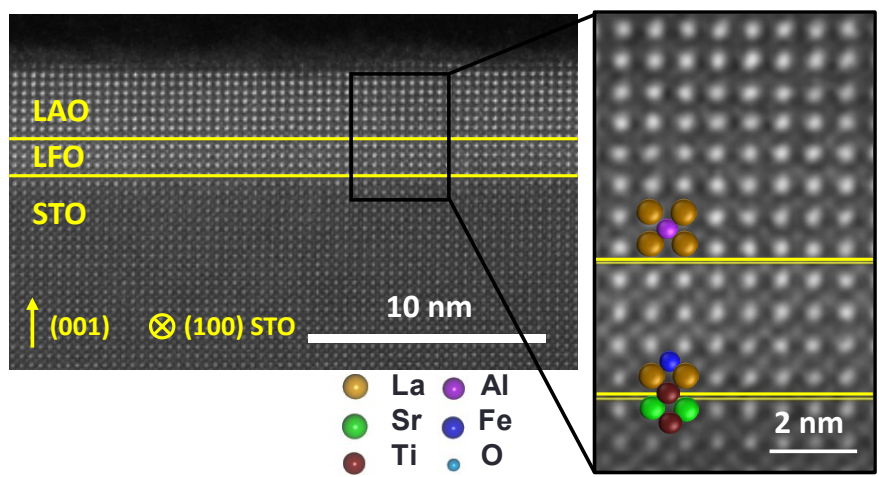

**Figure S4.** Annular dark-field scanning transmission electron microscopy (ADF-STEM) image of crystalline LAO/LFO(4 uc)/STO heterostructure.

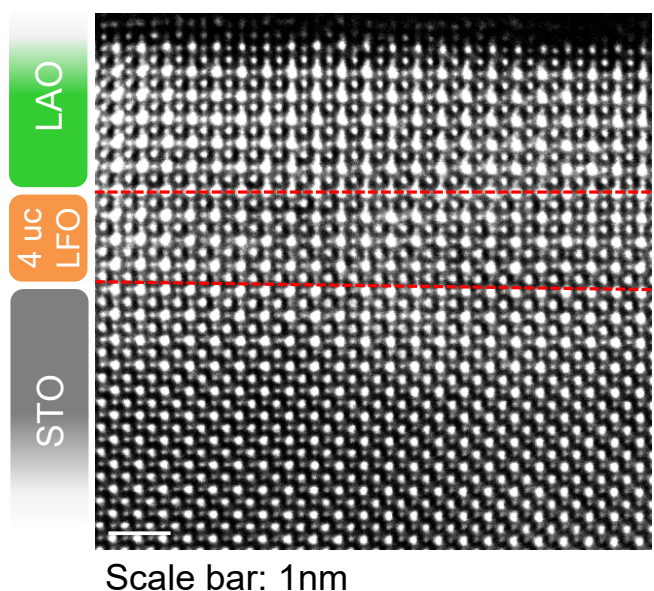

**Figure S5.** Annular bright-field scanning transmission electron microscopy (ABF-STEM). ABF-STEM results for LAO/LFO(4uc)/STO heterostructure. The ABF is more sensitive to the tilt of the electron beam with respect to the crystal zone axis than the ADF, and therefore oxygen atom columns can clearly be resolved.

The STEM images for LAO/LFO/STO heterostructures with $d = 0$, 2, and 4 uc also consistently present a coherent and epitaxial growth with sharp interfaces shown in Figure S4-S8. As discussed earlier, the charge transfer associated with the formation of $Ti^{3+}$-ions at the top layer of the STO results in a 2DES at the LAO/STO interface. In the presence of a LFO buffer layer, a fraction of the transferred electrons would be trapped in the LFO layer leading to the formation of $Fe^{2+}$-ions at the LAO/LFO interface. Analysing the EELS of the Fe-$L_{2,3}$ edges of the $d = 2$ uc sample confirms that 26.2% (1.2%) of the Fe atoms in the top (bottom) LFO layer indeed changes from $Fe^{3+}$ to $Fe^{2+}$ (Figure S6(b)). Likewise, as shown in Figure S6(c), the Ti-$L_{2,3}$ $e_g$ peaks confirm that 20.8% and 4.2% of Ti atoms at the topmost and second unit-cell layers of the STO change the valency from 4+ to 3+, while it remains 4+ for the rest of the STO layer underneath. The $Fe^{2+}$ and $Ti^{3+}$ concentration and distribution for the $d = 0$, 2 and 4 uc samples are presented in Figure S7-S9.

To extract the changes in the valency of cationic species, we analysed the energy shifts in the Ti-$L_{2,3}$ and Fe-$L_{2,3}$ edges. The Ti-$L_{2,3}$ $e_g$ peaks extracted from the interfacial layers shift toward lower binding energy, indicating a decreasing Ti valence from 4+ to 3+. Hence, the energy difference between $L_3$ $e_g$- $t_{2g}$ edges is used to calculate the $Ti^{3+}$ fraction relative to overall Ti species but is not apparent in the Fe-$L_{2,3}$ edges. So, a Voigt function fitting is performed to calculate the intensity ratio between $L_3$ and $L_2$, which validates the Fe valence change from $Fe^{3+}$ to $Fe^{2+}$. This indicates the presence of larger $Fe^{2+}$ and $Ti^{3+}$ fractions near to the LAO/LFO and the LFO/STO interfaces, respectively.

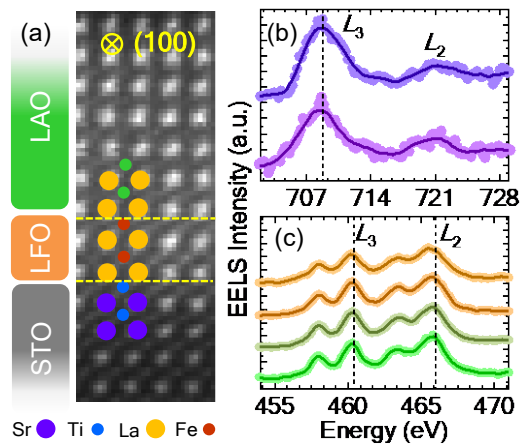

**Figure S6.** (a) An ADF image of LAO/LFO(2uc)/STO simultaneously acquired with EELS data. (b)-(c) Layer resolved Fe-$L$ and Ti-$L$ EELS at different $FeO_2$ and $TiO_2$ planes in the 2-uc-thick LFO and 4-uc-STO layers, respectively. The light-coloured dots are original data, and the dark-coloured solid lines are smoothed data by Fast Fourier Transform (FFT). The intensity ratio $L_3/L_2$ is used to extract $Fe^{2+}$ fraction and $L_3$ $e_g$-$t_{2g}$ is used to calculate $Ti^{3+}$ fraction.

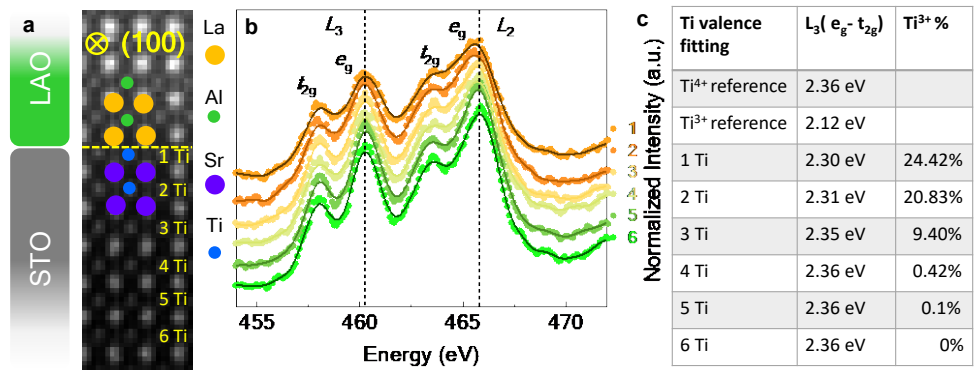

**Figure S7.** a, An ADF image of LAO/STO heterostructure simultaneously acquired with EELS data. b, Layer resolved EELS at different $TiO_2$ planes (numbered 1 through 6) in the STO substrate. A combination of the Gaussian and Lorentzian functions was used to fit the Ti- $L_{2,3}$ edge peaks and the results are summarized in the table. c, the sum of $Ti^{3+}$ fractions in the top 3-4 planes adds up about ~50% and reduces to 0% for the rest of the $TiO_2$ planes.

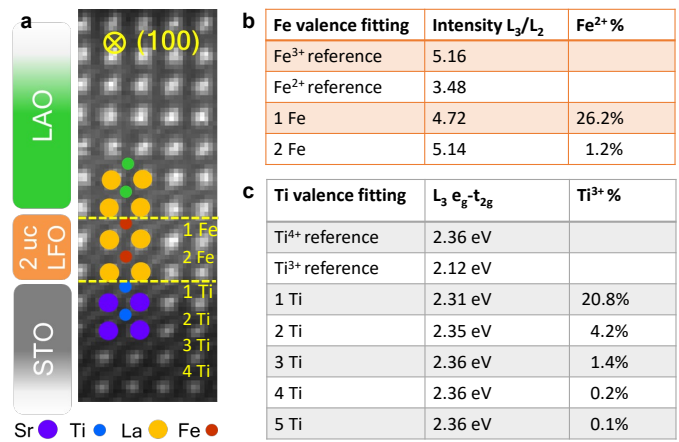

**Figure S8.** a, An ADF image of LAO/LFO(2 uc)/STO heterostructure simultaneously acquired with EELS data. A combination of Gaussian and Lorentzian functions was used to fit the Fe and Ti- $L_{2,3}$ edge peaks. The results are summarized in tables (b) and (c). b, $Fe^{2+}$ fraction is 26±5% in the top $FeO_2$ plane near the LAO/LFO interface (1 Fe) and reduces to nearly 0% for 2 Fe. c, the sum of $Ti^{3+}$ fractions in the top 2 planes are about ~25% and reduces to 0% for the rest of the $TiO_2$ planes.

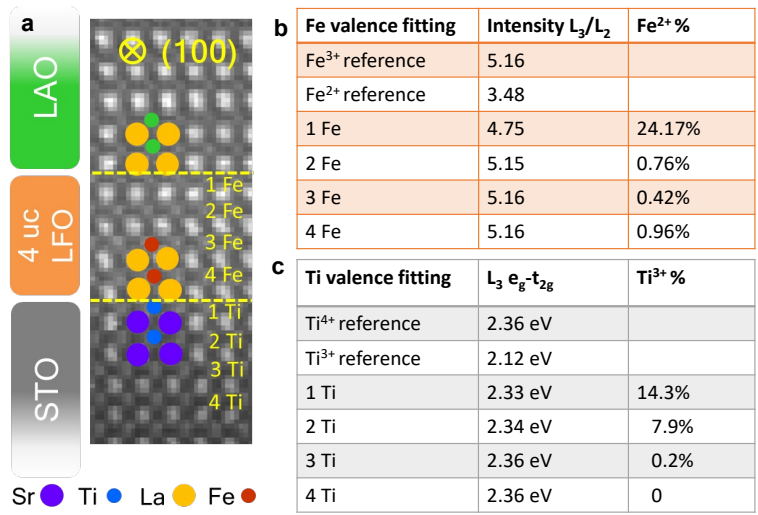

| Fe valence fitting | Intensity $L_3/L_2$ | $Fe^{2+}$ % |
|---|---|---|
| $Fe^{3+}$ reference | 5.16 | |
| $Fe^{2+}$ reference | 3.48 | |
| 1 Fe | 4.75 | 24.17% |
| 2 Fe | 5.15 | 0.76% |
| 3 Fe | 5.16 | 0.42% |
| 4 Fe | 5.16 | 0.96% |

| Ti valence fitting | $L_3$ $e_g$-$t_{2g}$ | $Ti^{3+}$ % |
|---|---|---|
| $Ti^{4+}$ reference | 2.36 eV | |
| $Ti^{3+}$ reference | 2.12 eV | |
| 1 Ti | 2.33 eV | 14.3% |
| 2 Ti | 2.34 eV | 7.9% |
| 3 Ti | 2.36 eV | 0.2% |
| 4 Ti | 2.36 eV | 0 |

**Figure S9.** a, An ADF image of LAO/LFO(4 uc)/STO heterostructure simultaneously acquired with EELS data. A combination of Gaussian and Lorentzian functions was used to fit the Fe and Ti- $L_{2,3}$ edge peaks. The results are summarized in tables (b) and (c). b, $Fe^{2+}$ fraction is 24±5% in the top $FeO_2$ plane near the LAO/LFO interface (1 Fe) and reduces to 0 for the rest of the $FeO_2$ planes. c, the sum of $Ti^{3+}$ fractions in the top 2 planes are about ~25% and reduces to 0% for the rest of the $TiO_2$ planes.

## SM6. Structural characteristics of 2DES interface.

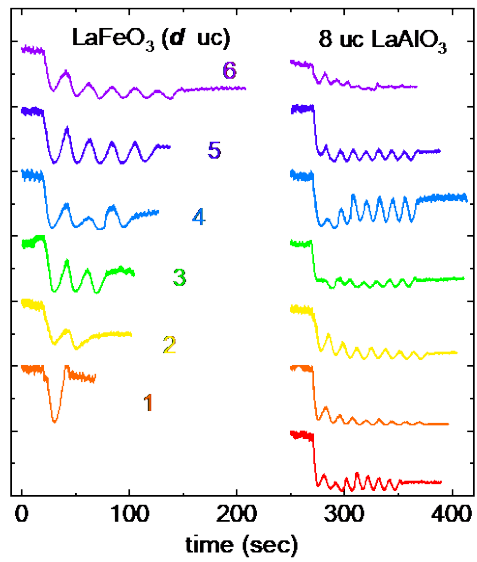

**Figure S10.** Layer-by-layer growth of LaFeO$_3$ with variable thickness ($d$ = 0 to 6 uc) and 8 uc LaAlO$_3$ on TiO$_2$ terminated SrTiO$_3$ interfaces determined by *in-situ* RHEED oscillations. We monitored the electrons reflection peak intensity along the (001) crystallographic direction during each growth, clearly signifies high-quality samples.

## SM7. Transport characterization of 2DES interface.

The all-electrical transport measurements (sheet resistance, carrier density and mobility) were carried in a quantum design physical property measurement system (PPMS) with temperature range of 300 down to 3 K and magnetic fields up to 9 T. To determine the exact carrier density, a Hall bar pattern on STO substrate was designed by a standard photolithography process. All lithography processes were performed (including AlN hard mask deposition) before growing the samples to avoid any exposure to the chemicals which can deteriorate the quality of surface and interface. The wires are connected by Al ultrasonic wire bonding for transverse and longitudinal transport measurement.

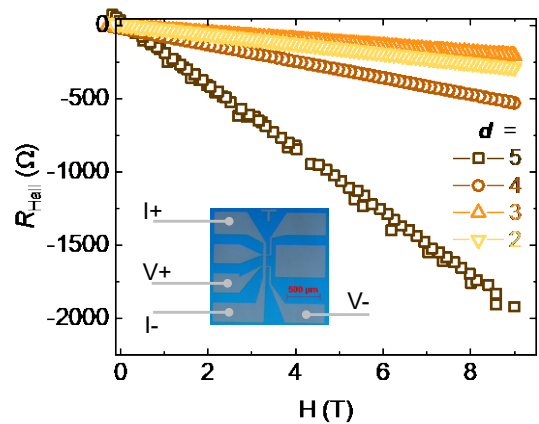

**Figure S11.** Hall resistance ($R_{Hall}$) versus magnetic field $H$ for different thickness ($d$) of LFO, A linear Hall effect is observed for all different thickness of LFO buffer layers. Inset is a Hall bar pattern on STO substrate was designed by a standard photolithography process.

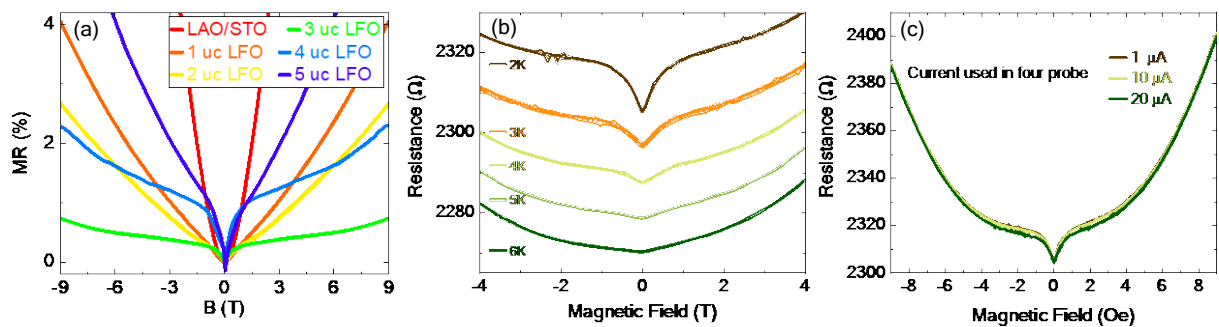

**Figure S12.** Anomalous Magnetoresistance. (a) Magnetoresistance curves for the various samples with different LFO thicknesses at temperature 3 K. (b) The magnetoresistance curves measured for 2 uc LFO spacer on LAO/STO structure as a function for temperature, between 2 K to 6 K. The minimum of the curve is determined mainly by the value of $H_{SO}$. From the plot, the minimum shifts with temperature, reflecting the decrease in the spin-orbit scattering field, (c) Magnetoresistance curves measured at different applied current values of 1, 10 and 20 $\mu A$.

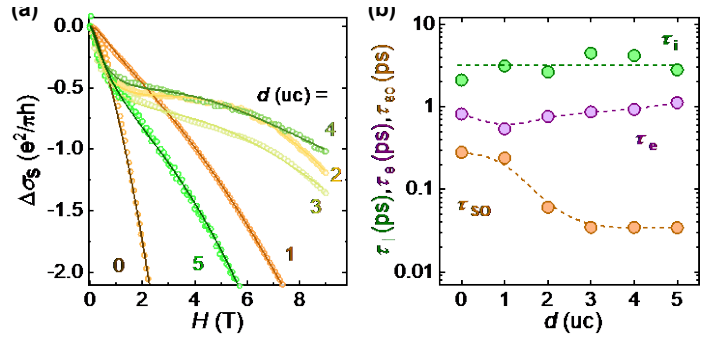

**Figure S13.** (a) Conductance ($\Delta\sigma_s$), normalized to $e^2/\pi h$, and best fits according to the Maekawa-Fukuyama theory for various $d$ values. Experimental data are shown in open circles, whereas theoretical fits are shown in solid lines, (b), Inelastic relaxation time $\tau_i$ (green circles), spin relaxation time $\tau_{so}$ (orange circles) and elastic relaxation $\tau_e$ (purple circles) as a function of $d$ plotted on a logarithmic time scale.

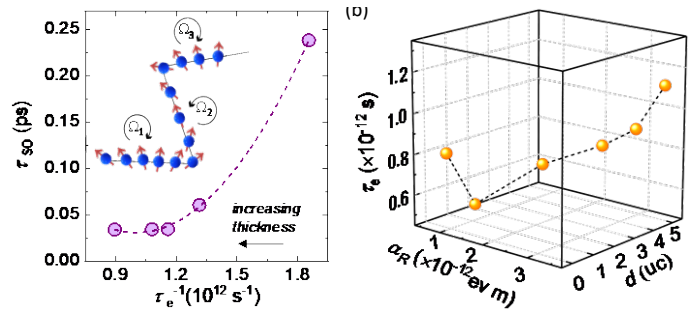

**Figure S14.** (a) Spin relaxation time ($\tau_{so}$) vs elastic scattering rate ($\tau_e$) dependency is consistent with D'yakonov-Perel' spin-relaxation mechanism (the electron spin precesses around the SOC fields and scatter on a timescale with a different vector of the SOC field corresponding Larmor frequency $\Omega_i$, inset figure). (b) scattering time (long carrier lifetime, $\tau_e$) and Rashba coefficients $\alpha_R$ plotted 3D as a function of the $d$.